\documentclass[prl,aps,twocolumn,epsfig]{revtex4}
\usepackage[dvips]{graphics,epsfig}

\def\beqn{\begin{equation}}
\def\eeqn{\end{equation}}
\def\beqnar{\begin{eqnarray}}
\def\eeqnar{\end{eqnarray}}
\def\ba{\begin{array}}
\def\ea{\end{array}}

\begin{document}

\newcommand{\ket}[1]{$\vert${#1}$\rangle$}
\newcommand{\mket}[1]{\vert{#1}\rangle}
\newcommand{\mbra}[1]{\langle{#1}\vert}
\newcommand{\tfrac}[2]{{\textstyle\frac{#1}{#2}}}
\def\sig1{\sigma_1}
\def\sig2{\sigma_2}
\def\sig3{\sigma_3}
\def\isig1{\iota\sigma_1}
\def\isig2{\iota\sigma_2}
\def\isig3{\iota\sigma_3}
\def\up{|\uparrow\,\rangle}
\def\dn{|\downarrow\,\rangle}
\def\upd{\langle\, \uparrow |}
\def\dnd{\langle\, \downarrow |}
\def\beqn{\begin{equation}}
\def\eeqn{\end{equation}}
\def\beqnar{\begin{eqnarray}}
\def\eeqnar{\end{eqnarray}}
\def\ba{\begin{array}}
\def\ea{\end{array}}

\title{A Method for Modeling Decoherence on a Quantum Information Processor}
\author{G. Teklemariam$^{\dagger}$, E. M. Fortunato$^{\ddagger}$, C. C. L\'opez$^{\S}$,
J. Emerson$^{\ddagger}$, J. P. Paz$^{\S}$, T. F. Havel$^{\ddagger}$ and D. G. Cory$^{\ddagger}$}

\affiliation{$^{\dagger}$Department of Physics, MIT \\
$^{\ddagger}$Department of Nuclear Engineering, MIT \\
$^{\S}$Departamento de F\'\i sica, ``Juan Jos\'e 
Giambiagi'', Facultad de Ciencias Exactas y Naturales, UBA}

\begin{abstract}
We develop and implement a method for modeling decoherence processes on an N-dimensional 
quantum system that requires only an $N^2$-dimensional quantum environment and 
random classical fields. 
This model offers the advantage that it may be implemented on small quantum information processors in order 
to explore the intermediate regime between semiclassical and fully quantum models. 
We consider in particular  
$\sigma_z\sigma_z$ system-environment couplings which induce coherence (phase) damping, though the model is directly extendable to other
coupling Hamiltonians.  Effective, irreversible phase-damping of the system is obtained by applying an 
additional stochastic Hamiltonian on the environment alone,  
periodically redressing it and thereby irreversibliy
randomizing the system phase information that has leaked into the environment as a result of the coupling.  This model is exactly
solvable in the case of phase-damping, and we use this solution to describe the model's behavior in some limiting
cases.  In the limit of small stochastic phase kicks the system's coherence decays exponentially at a rate which increases linearly 
with the kick frequency.  In the case of strong kicks we observe an effective decoupling of the system from the environment.  
We present a detailed implementation of the method on an nuclear magnetic resonance quantum information processor.  
\end{abstract}


\maketitle

\section{I. Introduction}
As early as the 1930s von Neumann \cite{vonNeumannBook} recognized that quantum correlations are crucial to understanding
the quantum measurement process. He considered measurement as a process that first required correlating the 
system with the quantum apparatus through a unitary, information conserving, quantum evolution. To complete the measurement
a mechanism was needed by which this pure, correlated state decayed into a mixture approximately diagonal in the basis 
of observation. In recent decades, the process of decoherence, which explains the dynamical origin of the above
decay, has been extensively studied \cite{ZurekPT,GiuliniBook,PazZurek,Zurek}. By employing an open-systems approach,
the effect of the interaction between the measurement apparatus and its environment was included explicitly, and von Neumann's method
was extended. The physical origin of the process of decoherence is very simple: the quantum correlations between
the apparatus and the environment that are established in the course of their interaction is responsible for the 
dynamical selection of a preffered set of states of the apparatus (the pointer states). 
The mechanisms of decoherence are now a subject of great practical interest. 
Some of the recent work on decoherence 
includes the determination of emergent properties of pointer states \cite{ZHP,PZ99}, efforts to design 
specific pointer states by engineering the environment \cite{Engeneering}, and identification of the time-scales
of the decoherence process \cite{DecoRate}. 

One of the simplest and most illustrative models of decoherence was originally suggested and studied by Zurek \cite{Zurek}. It
consists of a two level system (a spin $1/2$ particle) coupled to $n$ two level systems through a $\sigma_z\sigma_z$ type
interaction. With this model, in the large $n$ limit, it is possible to show that the correlations which arise between
the system and the environment lead to the damping of the system coherence, encoded in the off-diagonal elements of the
density matrix. In this work we will present results, both theoretical and experimental, for a two-level system that is coupled
to a few other two-level systems, which shows that by manipulating the latter one can reproduce the essential features of Zurek's model. 

Interest in this and other decoherence models (for example a two-level system coupled to a boson bath \cite{Ekert,Boson,Legget}), has
grown over the last few years due to the development of quantum information processing (QIP). A major challenge in QIP is the
preservation of quantum coherence in the face of constant perturbations by an environment. While one could try to isolate the QIP
device, this would make controlling the system difficult. Therefore, other strategies like quantum error correction (QEC)
\cite{Knill-QEC} and noiseless subsystems (NSs) \cite{Knill-NS,Viola-NS,Fortunato-DFS} have been developed. The aim of this work
is to develop methods to emulate decoherence in a physical setting, such as a QIP device, so that the nature and underlying
physics of decoherence can be better understood and applied in the development of control strategies.


The paper is organized as follows. In Section II we introduce the essential features of decoherence reviewing the model 
proposed by Zurek in which the system consists of a single spin while the environment is composed of an ensemble of $n$ spins. In
Section III we describe a simple model in which the environment is limited to only a few spins (qubits) and 
analyze a strategy through which these few spins can 
simulate a much larger effective 
environment. The strategy consists of randomly redressing the phase of the environment
qubits during their interaction with the system and averaging over many realizations of this evolution. 
We describe an exact solution of this model 
in the case of a $\sigma_z \sigma_z$ coupling between the system and a single environment qubit. 
In this case 
we provide an analytic description of the decoherence (phase-damping and decoupling) 
effects that arise under specific limiting conditions and also derive the 
associated Kraus operators for the model. A more detailed numerical analysis of this model 
is given for the case in which the environment consists of two qubits. 
In section IV, we present nuclear magnetic resonance (NMR) QIP simulations for the two qubit environment and comparisons
of these results with the one and two qubit environment predictions and numerical simulations. 
In Section V, we summarize our results and 
discuss the extension of this model to more general decoherence mechanisms.  

\section{II. Zurek's Decoherence Model}
In this section we review the basic elements of quantum decoherence by presenting an 
open-system model due to Zurek \cite{Zurek} which is simple enough to be solved analytically. In spite of its
simplicity the model captures many of the elements of decoherence theories and sheds insight into 
the loss of coherence, the onset of irreversibility,  and in particular, the role played by the size of the environment.

Consider $n$ two-level systems and focus on one system as the subsystem of interest. This subsystem interacts with the rest of the
system through a bilinear interaction. The 
overall dynamics is described by
\beqn
{\cal H}_{SE} = \sum^{n}_{k=2}J_{1k}\sigma^1_z\sigma^k_z,
\label{EQ3}
\eeqn
where the system qubit is denoted by the superscript '$1$'. This Hamiltonian is energy conserving and only causes phase damping. The
prescription of the open-systems approach is to evolve the combined system and environment, represented by the density matrix
$\rho^{SE}(t)$, and then recover the system density matrix from a partial trace over the environment degrees of freedom:
\beqn
\rho^S(t)=Tr_E\{\rho^{SE}(t)\}=\left( \begin{array}{cc} \rho^S_{00}(t) & \rho^S_{01}(t) \\
\rho^S_{10}(t) & \rho^S_{11}(t) \end{array} \right).
\label{RHOS00}
\eeqn
In Eq.~\ref{RHOS00} $\rho^S_{00}(t)$ and $\rho^S_{11}(t)$ represent the system population terms while $\rho^S_{01}(t)=\rho^{S*}_{10}(t)$
represents the system coherence term. 
If the coherence terms vanish, the pure state is turned into a mixture in the computational basis ($\sigma_z$-basis), i.e.
a ``pointer basis'' has been selected out by {\it einselection}. 
An important observation is that, in the absence of a self-Hamiltonian, 
the system's statonary states are selected out by the interaction Hamiltonian. 
In fact, since $[\sigma_z,{\cal H}_{tot}]=0$, the interaction
with the environment has two memory states $\mket{0}_S,\mket{1}_S$ as eigenstates and 
the populations remain unchanged throughout the system's evolution.
The coupling in Eq.~\ref{EQ3} is therefore a purely phase damping mechanism and
there is no energy exchange between system and environment. 

The combined system evolves by the unitary propagator
\beqn
{\cal U}_{SE}(t)= \exp(-i{\cal H}_{SE}t)=\exp(-i\sum^{n}_{k=2}J_{1k}\sigma^1_z\sigma^k_zt).
\eeqn
Consider a factorizable initial state of the combined system:
\beqnar
\mket{\Phi(0)}_{SE} &=& \mket{\psi(0)}_S\otimes\mket{\psi(0)}_E \nonumber \\
&=& (a\mket{0}_1+b\mket{1}_1)\prod^n_{k=2}(\alpha_k\mket{0}_k+\beta_k\mket{1}_k).
\eeqnar
The evolution is such that 
\beqnar
\mket{\Phi(t)}_{SE}
\hspace{0cm}=a\mket{0}_1\prod^n_{k=2}e^{-iJ_{1k}\sigma^k_zt}\mket{\phi}_k \hspace{2.19cm} \nonumber \\
+b\mket{1}_1\prod^n_{k=2}e^{iJ_{1k}\sigma^k_zt}\mket{\phi}_k,
\eeqnar
where $\mket{\phi}_k=\alpha_k\mket{0}_k+\beta_k\mket{1}_k$. The interaction
entangles the system states with the environment. In the language of QIP, 
the transformation ${\cal U}_{SE}$ generates a conditional phase gate between the system and its environment,
conditioned on the system's state. After the interaction the state is
\beqnar
\mket{\Phi(t)}_{SE}=a\mket{0}_1\prod^n_{k=2}[\alpha_ke^{-iJ_{1k}t}\mket{0}_k+\beta_ke^{iJ_{1k}t}\mket{1}_k]
\hspace{.35cm} \nonumber \\
+b\mket{1}_1\prod^n_{k=2}[\alpha_ke^{iJ_{1k}t}\mket{0}_k+\beta_ke^{-iJ_{1k}t}\mket{1}_k]
\eeqnar
and reflects the fact that the system and environment states are not factorizable. 
The off-diagonal element of the system's reduced density matrix (system coherence) is 
\beqn
\rho^S_{01}(t)={_1\mbra{0}}Tr_E\{\mket{\Phi(t)}_{SE}\mbra{\Phi(t)}_{SE}\}\mket{1}_1,
\eeqn
so that
\beqn
\rho^S_{01}(t)=ab^*z(t)
\eeqn
where
\beqn
z(t)=\prod^n_{k=2}[|\alpha_k|^2e^{-2iJ_{1k}t}+|\beta_k|^2e^{2iJ_{1k}t}].
\eeqn
Recall that $a$ and $b$ are the coefficients of the initial pure state of the system.

The time-dependence of $z(t)$ contains the crucial information for understanding the behavior of the system coherence. In
particular, the magnitude of $z(t)$ determines the damping of the phase information originally contained in
$\rho_{01}(0)$, with 
$|z(t)|\rightarrow0$ reflecting non-unitary evolution and ``irreversibility''. 

For a finite system, $|z(t)|$ is at worst quasi-periodic and one can always define  
a recurrence time, $\tau_E$. The existence of such a recurrence time reflects the fact that 
the information loss is in principle recoverable. 
In the continuum limit, $n\rightarrow\infty$,
$z(t)$ is no longer quasi-periodic and $\tau_E\rightarrow\infty$. 
The phase information is then unrecoverably lost, displaced from the degrees
of freedom of the system to the infintely many degrees of freedom of the environment.

To characterize the degree of decoherence 
one can consider the  size of the fluctuations of $z(t)$ around its time-averaged mean value $<z(t)>=0$:
\beqnar
<|z(t)|^2>
&=&
\lim_{T\to\infty}\frac{1}{T}\int^T_0dt'|z(t')|^2
\nonumber \\
&\approx&\frac{1}{2^{n-1}}\prod^n_{k=2}[1+(|\alpha_k|^2-|\beta_k|^2)^2]. \nonumber \\
\eeqnar
Thus, typical fluctuations vary as $\tfrac{1}{\sqrt{dim{\cal H}_E}}$ and 
the effectiveness of the decoherence mechanism in this model is determined by  
the dimension of the environment.

To summarize, the key features of this model of decoherence 
are: (1) the system of interest
evolves through a direct entangling interaction with each two-level system in a very large environment, 
and at any time the (reduced) density matrix of the system is obtained from a trace over 
the environment degrees of freedom; 
(2) expressed in the pointer basis of the system, which in this simplest case  
is the set of states that commute with the interaction Hamiltonian, 
the reduced density matrix becomes approximately diagonal and the off-diagonal 
elements exhibit coherence loss; (3) the 
fluctuations of the decoherence produced by this model, measured by the size of 
the system's off-diagonal elements, are   
controlled by the dimension of the environment's Hilbert space. 

\section{III. A Hierarchical Decoherence Model}

\begin{figure}
	\begin{center}
  {\epsfxsize=3.5in\epsfysize=2.5in\centerline{\epsffile{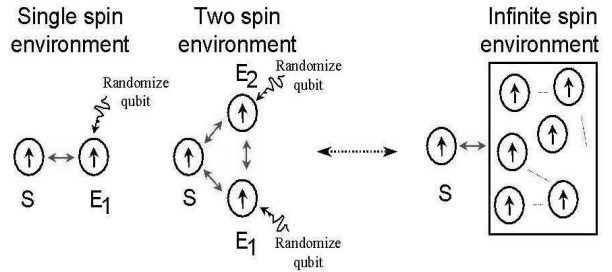}}}
\vspace{.0cm}
			\caption {The schematics on the left describe the models developed in this paper. 
				A single spin system, $S$, is coupled to one and two spin quantum environments, 
				designated $E_i$. 
				During the coupling the system
				phase information leaks into the environment. Since the spin environment is finite, in order 
				to simulate the effects of a larger quantum environment (depicted on the right) a mechanism is
				needed by which the information stored in the available quantum environment can be effectively erased. 
				We accomplish this by 
				redressing the environment
				degrees-of-freedom with stochastic phase kicks. 
				\label{DecoFig0} }
	\end{center}
\end{figure}

\begin{figure}
	\begin{center}
  {\epsfxsize=3.5in\epsfysize=2.85in\centerline{\epsffile{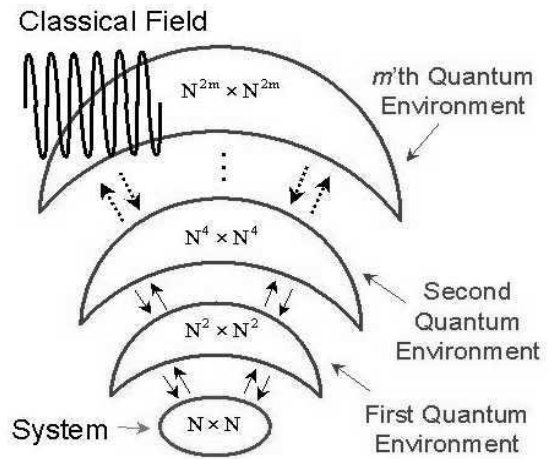}}}
			\caption {A schematic of the system coupled to a hierarchy of quantum environments, as resources 
				permit, and
				the  role of the classical stochastic field. 
				Each environment is coupled only locally (to its neighbours in the hierarchy). 
				It should be noted  that the classical environment 
				only interacts with the quantum environments, and does not interact directly 
				with the quantum system. In this paper we consider only the case of one 
				local quantum environment, as portrayed in Fig.~1. \label{DecoFig1} }
	\end{center}
\end{figure}

We consider the problem of experimentally simulating quantum decoherence in a physical setting in which 
limited quantum resources are available for modeling the quantum environment.  
By ``simulating quantum decoherence'' we are referring not only to the challenge of implementing 
an arbitrary open-system trajectory on a QIP device, but also to the study of the decoherence processes 
that result from specific system-environment couplings (for example, derived from a model of  
some physical system of interest). 

As in Zurek's model the exclusive direct mechanism for system decoherence in our model 
is the coupling between the system and a local quantum environment 
through a fixed bilinear Hamiltonian. 
However our model of decoherence has two distinct features from the model described above. 
The first difference is a constraint on the Hilbert space size derived from practical considerations: 
we allow the dimension of the Hilbert space for 
the local quantum environment to be no larger than $N^2$, where $N$ is the 
dimension of the Hilbert space of the system. 
In this way the quantum environment is the smallest size that will enable the implementation 
of an arbitary completely positive map on the system through a unitary operator 
on the combined system and environment.  
To remove the information from our finite quantum environment we include a stochastic classical field in our model. 
This strategy is designed to eliminate the quantum back-action from low dimensional 
environments. Basically the technique consists of 
redressing \cite{CC-T} the environment's quantum state by applying a sequence of random classical kicks 
to the environment qubits, and then averaging over realizations of this stochastic noise. 
This has the effect of scrambling the system information after it has been stored in the quantum environment through 
the coupling interaction. 
It is worth stressing that the system itself is not subjected to these classical kicks and the 
associated stochastic averaging. 
This model, and the associated method realized in this paper, is 
depicted schematically in Fig.~1. A generalization of this method to provide a time-dependent 
open-system evolution, is described in the discussion 
and depicted in Fig.~2. 

As we shall show below, this scheme enables simulation of the quantum decoherence 
that normally arises for much larger effective environment sizes.
In particular, we demonstrate the 
simulation of phase-damping on an NMR QIP consisting of three qubits (see Fig.~1). 
In the NMR simulation, the system is represented by one qubit while the other two qubits represent the 
quantum environment. Before turning to a discussion of the three-qubit experiment, 
we first describe and analyze this simulation method theoretically 
in the simplest and solvable case of the phase-damping of a single system qubit 
from a single environment qubit (also depicted in Fig.~1).

\subsection{One-qubit Environment: Simple Solvable Model}

Below we introduce the essential features of this decoherence model by considering 
an exact solution available in the case of a one-qubit environment coupled to the system by a $\sigma_z \sigma_z$ interaction. 
With the system and environment qubits labeled by $S$ and $E$, respectively, the full Hamiltonian 
is given by
\begin{equation}
{\cal H}_{0}=\pi(\nu_S\sigma^S_z+\nu_E\sigma^E_z+\frac{\Omega}{2}\sigma^S_z\sigma^E_z).
\label{Hint}
\end{equation}
Here, $\nu_S$, $\nu_E$ and $\Omega$ are frequencies in units of Hertz. This Hamiltonian includes both the self-evolution of the two
qubits and their interaction. In the absence of any other interaction, the evolution operator for a time $t$ is
\begin{equation}
U(t)=\exp\left[-i\pi (\nu_S\sigma^S_z+\nu_E\sigma^E_z+\frac{\Omega}{2}\sigma^S_z\sigma^E_z ) \; t\right].
\label{U0}
\end{equation}
We will consider the evolution of this system subject to a sequence of kicks that affect only the environment qubit. 
Every kick is generated
by a transverse magnetic field that rotates the environment qubit around the $y$-axis by an angle $\epsilon_m$ chosen randomly in the
interval $(-\alpha,+\alpha)$. The evolution operator for the $m$-th kick is given by $K_1 = I^S \otimes K_1^E$, where  
\begin{equation}
K_m^E=\exp(-i\epsilon_m \sigma^E_y).
\end{equation}
and $I$ is the identity matrix.
In our proposed model, the kicks are considered instantaneous, therefore 
the evolution for a total time $T=n/\Gamma$,  
where $\Gamma$ is the kick rate, can be written,  
\begin{equation}
U_{n}(T) = K_{n} U\left(\frac{T}{n}\right)K_{n-1} U\left(\frac{T}{n}\right)\ldots K_1 U\left(\frac{T}{n}\right).\label{Un}
\label{evol}
\end{equation}
It is useful to keep in mind that the operator $U_{n}(T)$ 
depends also on the values of the random variables $\epsilon_m$ ($m=1,...,n$) corresponding to the kick angles. 

Our goal is to obtain a closed expression for the reduced density matrix of the system qubit for an ensemble of
realizations of the random variables $\epsilon_m$. The density matrix for this ensemble is given by
\begin{equation}
\overline{\rho^S(T)}=\int_{-\alpha}^{\alpha}{d\epsilon_n\over 2\alpha}\ldots
\int_{-\alpha}^{\alpha}{d\epsilon_1\over 2\alpha} Tr_E\left[U_n \rho^{SE}(0) U_n^\dagger\right],
\label{rho1big}
\end{equation}
where $\alpha$ is the spread of allowed kick angles over which the $\epsilon_m$ ($m=1,...,n$) are 
uniformly distributed. We will consider a factorizable
initial state for the two qubits (this is not essential):
\begin{equation}
\rho^{SE}(0)=\rho^S(0)\otimes\rho^E(0).\label{init12}
\end{equation}
It is convenient to express the initial density matrix of the system in the basis of eigenstates of $\sigma_z$, 
\begin{equation}
\rho^S(0)=\sum_{j,l=0,1}\rho^S_{jl}(0)|j\rangle\langle l|.\label{init1}
\end{equation}
Then we can simplify the expression for $U_n(T)$. 
To do this, we evaluate the effect of the first step in the evolution Eq.~\ref{evol} as follows,
\begin{eqnarray}
\rho^{SE}(1) & = &   
K_1 U(T/n) \; \rho^{SE}(0) \; U(T/n)^\dagger K_1^\dagger  \\
& = & \sum_{j,l=0,1} 
 \left[ \rho^S_{jl}(0) \; |j\rangle\langle l| \right]  
\nonumber \\
& & \otimes \left[
K_1^E V_j^E
\rho^E(0)
(K_1^E V_l^E)^\dagger \right] \nonumber,
\end{eqnarray}
where we have defined the environment operator, 
\begin{eqnarray}
V_j^E & = & \langle j | U\left(\frac{T}{n}\right)| j \rangle  \\
& = & 
\exp\left[ -i \frac{\pi}{\Gamma}  \nu_S (-1)^j 
-i\frac{\pi}{\Gamma}\left(\frac{\Omega}{2}(-1)^j+\nu_E \right) \sigma^E_z \right]. \nonumber
\end{eqnarray}
In the above we have explicitly evaluated the action of the interaction Hamiltonian on the system states,  
and the $j$-dependence of the single-step operator $V_j^E$ 
reflects the fact that it operates on the environment state conditionally on the system state.
The important point is that 
the evolution operators for the additional $n-1$ iterations will factor as above, 
producing a final expression with (conditional) operators that act exclusively on the environment qubit. 
Hence, we can immediately obtain the following simple form for the final density matrix of the system qubit, 
\begin{equation}
\overline{\rho^S(T)}=\sum_{j,l=0,1}\rho_{jl}(0) f_{jl}(n,T)\ |j\rangle\langle l|,
\end{equation}
where the function $f_{jl}(n,T)$, which we call the {\em decoherence factor}, 
carries all the information about the effect of the environment qubit
on the system qubit, including also the trivial phases from the system's self-evolution. 
It is given by the formula
\begin{equation}
f_{jl}(n,T)=\int_{-\alpha}^{\alpha}\!{d\epsilon_n\over 2\alpha}\ldots
\int_{-\alpha}^{\alpha}\! {d\epsilon_1\over 2\alpha} 
Tr_E \! \left[ (A_j^E)_n \; \rho^E(0) \; \left(A_l^E\right)_n^\dagger \right],
\label{fjl}
\end{equation}
where the operator $(A_j^E)_nn$ is defined as
\begin{equation}
(A_j^E)_n = \langle j | U_n(T) | j \rangle = K_n^E V_j^E K_{n-1}^E V_j^E
\ldots K_1 V_j^E.
\label{Unj2}
\end{equation}
It is clear from Eq.~\ref{fjl} that for $j=l$ the final trace over the environment system is equal to  one and therefore we always have
$f_{jj}=1$. Thus, this decoherence model only affects the off-diagonal terms in the $\sigma_z$--basis, in other words, the
$\sigma_z$-eigenbasis is a pointer basis. 

The remaining task is to evaluate the decoherence factor $f_{01}(n,T)$ since on general grounds one can show that
$f_{jl}(n,T)=f_{lj}(n,T)^*$. 
To evaluate $f_{jl}(n,T)$ it is convenient to notice that the integrals in Eq.~\ref{fjl} can be brought forward through 
the independent operator terms in the sequence in Eq.~\ref{Unj2}, and the evolution can be expressed as 
the succesive application of a superoperator on the initial density matrix of the environment
$\rho^E(0)$. Thus, we can write:
\begin{equation}
f_{01}(n,T)=Tr_E\left[O^n(\rho^E(0))\right]
\end{equation}
where the superoperator $O$ is defined as
\begin{eqnarray}
O(\rho)&=&\int_{-\alpha}^{\alpha}{d\epsilon\over 2\alpha} 
K^E V_0^E \rho \; (V_1^E)^\dagger (K^E)^\dagger \nonumber \\
&=&e^{-2i\pi\nu_S T/n}\int_{-\alpha}^{\alpha}{d\epsilon\over 2\alpha} 
e^{-i\epsilon \sigma_y}e^{-i\pi(\frac{\Omega}{2}+\nu_E) T \sigma_z/n}
\nonumber\\
&\ &\times \rho \; e^{-i\pi(\frac{\Omega}{2}-\nu_E) T \sigma_z/n} e^{i\epsilon \sigma_y}.
\label{Orho}\end{eqnarray}
The dependence of $f_{01}$ on the self-evolution of the system factors out as a phase factor that modulates the overall evolution in 
Eq.~\ref{Orho}. This trivial phase factor will be omitted from here on because it can be easily restored if necessary. After integrating 
over the random variable the last expression becomes  
\begin{eqnarray}
O(\rho)&=& c \left(e^{-i\pi(\frac{\Omega}{2}+\nu_E) T \sigma_z/n}
\rho e^{-i\pi(\frac{\Omega}{2}-\nu_E) T /n}\right)+\\
       &\ & d\left(\sigma_y e^{-i\pi(\frac{\Omega}{2}+\nu_E) T \sigma_z/n}
\rho e^{-i\pi(\frac{\Omega}{2}-\nu_E) T \sigma_z /n} \sigma_y\right)\nonumber
\end{eqnarray}
where $\gamma = c-d=\sin(2\alpha)/2\alpha$ and $c+d=1$. It is worth stressing that this superoperator is 
not trace preserving or hermitian. 
It is easy to show that $O(\sigma_x)=\sigma_x$ and $O(\sigma_y)=\gamma\sigma_y$. Thus, both $\sigma_x$ and $\sigma_y$ are eigenvectors
of the superoperator $O$ (respectively with eigenvalue $1$ and $\gamma$). We will later need to find the other two eigenvectors which
are linear combinations of the identity $(I)$ and $\sigma_z$. The decoherence factor $f_{01}$ is given following a final trace
over the environment qubit. So, the traceless terms in $\rho^E(0)$, those proportional to $\sigma_x$ and $\sigma_y$ do not contribute to the
final result. Thus, to compute $f_{01}(n,T)$ the superoperator $O$ is applied $n$--times to the part of the initial state with components
along the identity and $\sigma_z$. Writing the initial density matrix of the environment qubit as
$\rho^E(0)=(I+p_x \sigma_x+p_y \sigma_y +p_z \sigma_z)/2$ we obtain
\begin{equation}
f_{01}(n,T)={1\over 2} Tr(O^n(I)) + {1\over 2} p_z Tr(O^n(\sigma_z)). 
\end{equation}
The action of $O$ on the identity and $\sigma_z$ is
\begin{eqnarray}
O(I)&=&\cos(\pi\Omega T/n) I-i \gamma \sin(\pi\Omega T/n) \sigma_z \nonumber \\
O(\sigma_z)&=&-i\sin(\pi\Omega T/n) I + \gamma \cos(\pi\Omega T/n) \nonumber
\sigma_z.
\end{eqnarray}
Note that the above expressions have no dependence on the frequencies $\nu_S$ and $\nu_E$ since they came in as trivial phase factors. 

The 
eigenvalues $\lambda_1$ and $\lambda_2$ (and the corresponding eigenvectors) of the superoperator $O$ 
can be obtained directly, giving,  
\begin{equation}
\lambda_{1\atop 2}={1\over 2} (1+\gamma)\cos(\pi\Omega T/n)
\pm\sqrt{{(1+\gamma)^2\over 4}\cos^2(\pi\Omega T/n)-\gamma} \label{lambda12} \nonumber
\end{equation}
and, from them one can find the following exact solution, 
\begin{eqnarray}
f_{01}(n,T) &=& {{\cos(\pi\Omega T/n)(\lambda_1^n-\lambda_2^n)+\lambda_1\lambda_2^n
-\lambda_2\lambda_1^n}\over (\lambda_1-\lambda_2)}\nonumber\\
&-&ip_z \sin(\pi\Omega T/n){(\lambda_1^n-\lambda_2^n)\over
(\lambda_1-\lambda_2)}.
\label{f01lambda}
\end{eqnarray}
Notice that this formula is an explicit expression (obtained with no approximations) valid for all values of the parameters defining our
model ($n$, $\gamma$, etc). Also, it is worth stressing that the dependence on the initial state of the environment (entering the above
equation through the initial polarization $p_z$) is rather trivial. Moreover, the first and second lines of the last equation clearly
separate the real and imaginary parts of the decoherence factor $f_{01}$. Below, we will analyze the predictions of this model for some
simple cases.

\begin{center}
{\bf Dependence on Kick Angle: Limiting Cases}
\end{center}

We first consider the dependence of the decoherence factor $f_{01}$ on $\gamma$. 
Let us consider three cases. First we discuss the limit $\gamma = 1$ that corresponds to unitary evolution (that is, 
no kicks since the kick angle $\alpha = 0$). 
Then, we consider the case $\gamma=0$ that corresponds to averaging over angles between $0$ and $2\pi$. Finally,
we analyze in some detail the case where $\gamma$ is close to one (small angle kicks), 
which is the condition met in our simulations and
experiments. In all these cases the decoherence factor $f_{01}$ is directly related to observable quantities 
$\langle\sigma^S_x\rangle=2\Re[\rho_{01}f_{01}]$ and $\langle\sigma^S_y\rangle=2\Im[\rho_{01}f_{01}]$.

\begin{center}
{\it Unitary evolution: $\gamma=1$}
\end{center}

This is the simplest case. Here, the superoperator $O$ is such that $O(\rho)=\rho \exp(-i\pi\Omega T \sigma_z/n)$ for any operator
$\rho$ that is a linear combination of the identity and $\sigma_z$. (Note that we showed earlier that $\sigma_x$ and $\sigma_y$ are
eigenvectors of $O$ and thus vanish after the trace). Using this, or simply replacing $\gamma=1$ in the
decoherence factor (Eq.~\ref{f01lambda}),
\begin{equation}
f_{01}(n=0,T)= \cos(\pi\Omega T) - i\;p_z\;\sin(\pi\Omega T).
\label{f01g1}
\end{equation}
This has a clear physical interpretation. The decoherence factor is independent of the kicking rate (as it should be since there are no
kicks in this limit). Recall that $p_z$ is the initial polarization of the environment qubit; therefore the system qubit rotates independently
of the environment qubit.

\begin{center}
{\it Complete randomization: $\gamma=0$}
\end{center}

Here the kick angles $\epsilon_j$ vary over the entire interval between $0$ and $2\pi$. In this case the above formulae
simplify substantially to
\begin{eqnarray}
f_{01}(\Gamma,T)&=&\cos^{\Gamma T}({\pi\Omega\over\Gamma}) \nonumber \\
&-&i\;p_z\;\sin({\pi\Omega\over\Gamma}) 
\cos^{\Gamma T-1}({\pi\Omega\over\Gamma}),
\label{f01g0}
\end{eqnarray}
where use of $n= \Gamma T$ was made. (Recall that $\Gamma =n/T$ is the kick rate). In the large $\Gamma$ limit we clearly see a
Zeno--like effect (for an operator not a state) that can be obtained from Eq.~\ref{f01g0} by noting that
\begin{equation}
\cos^{\Gamma T}({\pi\Omega \over \Gamma}) \approx 
(1-\frac{1}{2}({\pi\Omega \over \Gamma})^2)^{{\Gamma T}} 
\approx \exp(-{(\pi\Omega)^2T \over 2\Gamma}).
\label{Zeno}
\end{equation}
Thus, in this limit for faster kick rates $\Gamma$ the system takes longer to decohere.

\begin{figure}
	\begin{center}
\rotatebox{270}{{\epsfxsize=2.85in\epsfysize=3.5in\centerline{\epsffile{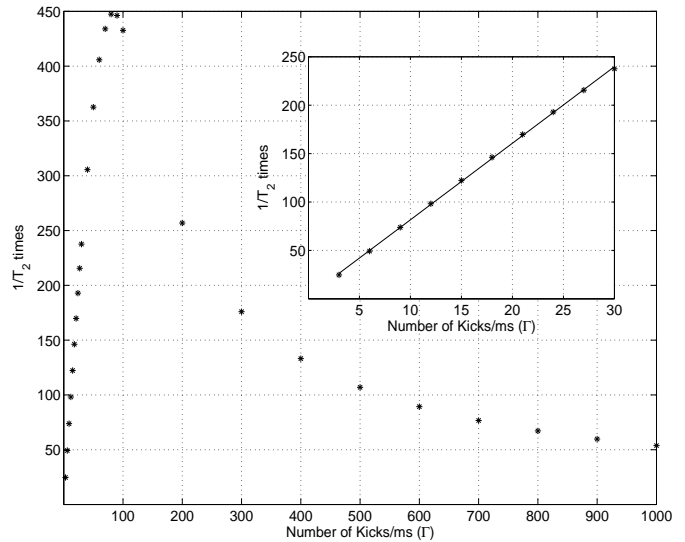}}}}
\vspace{.25cm}
			\caption{
The decay rate as a function of the kick rate $\Gamma$. 
For $\Omega=300Hz$ and $\gamma=0.98$ the kicking is no longer effective at 
 inducing decoherence beyond a kick rate of about $50$ kicks/ms. Only kick rates upto $1000$
kicks/ms are shown and after $5000$
kicks/ms the decay is no longer exponential.  
Inset: The decay rate as a function of the kick rate is linear
for small values of $\Gamma$. The plot is for $\Omega=300Hz$
and $\gamma=0.98$.  \label{LopezFig} }
	\end{center}
\end{figure}

\begin{center}
{\it Average over small angles: $\gamma=1-{\cal O}(\alpha^2)$}
\end{center}

Here we consider the case where the averaging is over small angles (the regime we consider in the simulations and experiments is
$\alpha=\pi/20$), where
\begin{equation}
\gamma \approx 1 - {2 \over 3}\alpha^2.
\label{epsilon}
\end{equation}
Defining $\epsilon = {2\over3}\alpha^2$ we can expand both eigenvalues in powers of $\epsilon$ to obtain an expression which is valid for
small $n=\Gamma T$:
\begin{eqnarray}
f_{01}(\Gamma,T)= \hspace{6.25cm}\nonumber \\ (1-{\epsilon \over 2})^{\Gamma T} 
(1+{\epsilon \over 2})[\cos(\pi\Omega T) - i\ p_z \sin(\pi\Omega T) + 
{\cal O}(\epsilon)]. \nonumber \\
\label{f01gnear1}
\end{eqnarray}
In this regime the envelope of the decay of $f_{01}$ is exponential 
with a decay rate proportional to the kick rate because $\epsilon \ll 1$
implies $(1-{\epsilon \over 2})^n \approx \exp(-n\epsilon)$. The analysis of the exact formula shows that in this case (large $n$) a
Zeno--type effect arises (as before). 

The dependence of the decay rate, $T_2$, as a function of the kicking rate is shown in
Fig.~\ref{LopezFig}. 
The numerical data in the figures are obtained from the exact expression for $f_{01}$.  
The initial state for the system qubit is taken to be $\rho^S=\tfrac{1}{2}(I+\sigma_x)$, in which case $f_{01}$ is directly
proportional to the transverse polarization of the qubit. For small values of the
kick rate $1/T_2$ is linear in $\Gamma$. However for larger values $1/T_2$ saturates and decays again due to the Zeno-like effect.
These results substantiate our expectation that the kick-rate can be applied to control the attenuation of the 
recurrences.  In the low kick-rate limit the role of the kick-rate is analogous 
to the variable environment size in Zurek's model. 
 



\begin{center}
{\bf Kraus Forms}
\end{center}

For a one-spin environment a phase damping channel can be represented by a purification basis \cite{PreskillNotes,Schumacher}
that evolves the system and environment with the unitary operator:
\begin{equation}
{\cal U}_{SE}=e^{-i\theta\sigma^S_z\sigma^E_y}=E^S_+e^{-i\theta\sigma^E_y/2}
+ E^S_- e^{i\theta\sigma^E_y/2},
\end{equation}
where $E^S_\pm=\frac{1}{2}(I\pm\sigma^S_z)$, 
or equivalently $E_+=\mket{0}\mbra{0}$ and $E_-=\mket{1}\mbra{1}$. 
This operator transforms the states of $\rho^{SE}(0)=\rho^S(0)\otimes E^E_+$ as follows:
\beqnar
\mket{0}_S\mket{0}_E \stackrel{\cal U}{\longrightarrow} \cos(\theta/2) \mket{0}_S\mket{0}_E
+\sin(\theta/2) \mket{0}_S\mket{1}_E, \nonumber \\
\mket{1}_S\mket{0}_E \stackrel{\cal U}{\longrightarrow} \cos(\theta/2) \mket{1}_S\mket{0}_E
-\sin(\theta/2) \mket{1}_S\mket{1}_E.
\eeqnar
By tracing away the environment states (${\cal H}_E:\{\mket{0}_E,\mket{1}_E\}$) this channel has the Kraus operator
sum representation \cite{Kraus,NielsenBook,Havel} given by
\beqn
\hat{\cal S}(\rho^S)=\hat{M}_0\rho^S\hat{M}_0+\hat{M}_1\rho^S\hat{M}_1,
\eeqn
where
\beqn
\hat{M}_0=\cos(\theta/2)I^S,\;\;\;\hat{M}_1=\sin(\theta/2)\sigma^S_z,
\eeqn
and $\hat{\cal S}$ is the superoperator map,
\beqn
\hat{\cal S}(\rho^S)=\left[ \begin{array}{cc} \rho^S_{00} & \beta \rho^S_{01} \\
\beta \rho^S_{10} & \rho^S_{11}
\end{array} \right],
\eeqn
with $\beta=\cos^2(\theta/2)-\sin^2(\theta/2)$. If we parametrize
\beqn
\cos(\theta/2) \equiv \sqrt{\frac{1}{2}(1+f_{01})}, \;\;\; \sin(\theta/2) \equiv \sqrt{\frac{1}{2}(1-f_{01})},
\eeqn
then we obtain the Kraus operator sum representation for the phase damping channel in our model:
\beqn
\hat{\cal S}(\rho^S)=\frac{1}{2}(1+f_{01})\rho^S+\frac{1}{2}(1-f_{01})\sigma^S_z\rho^S\sigma^S_z.
\eeqn



From the analytical solution to the two-qubit model 
we see that a single qubit environment interacting with a single qubit system 
is sufficient to represent the phase-damping channel. Similarly, an $N$-dimensional 
system interacting with an environment of dimension $N$ 
through the $\sigma_z \sigma_z$ interaction 
is sufficient to describe the open-system dynamics 
of phase-damping. 
This is because dephasing is a special case where the Lie algebra 
of the noise consists of only the two operators $\sigma_z$ and $I$ (out of a possible four). 
In contrast, for an arbitrary completely positive map 
the dimension of the environment must be at least $N^2$ for a system  
with dimension $N$ to induce an arbitrary  mapping on the system.

\subsection{Two-Qubit Environment: Numerical Simulation}

In the more general case where we wish to implement any completely positive map \cite{Kraus,NielsenBook} on 
one-qubit system the minimum required environment is two qubits. 
We therefore want to consider a two-qubit environment model. Moreover, 
we want to examine the effect of only a finite number of realization of the random 
kick variables. Therefore, a three-qubit model is explored numerically below. 
The results of on an NMR QIP simulation \cite{Somaroo,Tseng} of this model are presented in the next section.

\begin{table}
	\begin{center}
		\setlength{\unitlength}{.35cm}
		\begin{picture}(10,10.5)

		\put(0,9.5){$\nu_S=0$}
		\put(0,8.0){$\nu_{E_1}=630 Hz$}
		\put(0,6.5){$\nu_{E_2}=-630 Hz$}
		\put(0,5.0){$J_{SE_1}=250 Hz$}
		\put(0,3.5){$J_{SE_2}=50 Hz$}
		\put(0,2.0){$J_{E_1E_2}=174 Hz$}
		\put(0,0.5){$\theta=\{-\tfrac{\pi}{20},+\tfrac{\pi}{20}\}$ (randomly choosen)}

		\end{picture}
	\end{center}
		\caption{This table lists the parameters for the model Hamiltonian of
				Eq.~\ref{SysHam}. \label{Tab1} }
\end{table}

We now consider the following system-environment Hamiltonian, 
\beqn
{\cal H}_{tot}={\cal H}_S+{\cal H}_E+{\cal H}_{SE}+{\cal H}_{E_1E_2}, 
\label{SysHam}
\eeqn
where, 
\begin{eqnarray}
{\cal H}_S  & =&  \pi\nu_S\sigma^S_z, \nonumber \\
{\cal H}_E & = & \pi\sum^2_{i=1}\nu_{E_i}\sigma^{E_i}_z, \nonumber \\
{\cal H}_{SE} & = & \frac{\pi}{2}\sum^2_{i=1}J_{SEi}\sigma^S_z\sigma^{Ei}_z, \nonumber \\
{\cal H}_{E_1E_2} & = & \frac{\pi}{2}J_{E_1E_2}\sum_{i=x,y,z}\sigma^{E_1}_i\sigma^{E_2}_i, \nonumber 
\end{eqnarray}
The environment spins, $E_1$ and $E_2$, are also subjected to periodic, instantaneous kicks with an evolution 
operator of the form, 
\beqn
{\cal K}_m = \exp\left[ \sum^2_{i=1}\theta_i^m\sigma^{Ei}_y \right], 
\label{KickHam}
\eeqn
where the $\theta_i^m$ are the random values of the $m$'th kick.
The instantaneous nature of the kicks allows the evolution of the full system over the time interval $T$, with $n$ instaneous 
kicks, to be described by the operator,  
\begin{eqnarray}
{\cal U}_n  & = &  
{\cal K}_n \exp\left[-i{\cal H}_{tot} (T/n) \right] {\cal K}_{n-1} \exp\left[-i{\cal H}_{tot} (T/n) \right] \nonumber \\
& & 
\times \cdots \times \; {\cal K}_1 \exp\left[-i{\cal H}_{tot} (T/n) \right], 
\end{eqnarray}
where 
each ${\cal K}_m$ has a different random kick variable.

The resultant system density matrix for a single realization is now obtained by tracing out the environment.  
As before, we are interested in the system coherence as expressed through the off-diagonal elements 
of the system state in the basis of the pointer states, 
\beqn
\mbra{0} \rho^S(T) \mket{1} =  \mbra{0}Tr_{E_1E_2}\left[{\cal U}_n\rho^{SE_1E_2}(0) {\cal U}_n^\dagger \right]\mket{1}_S ,
\eeqn
Finally, we must average over different realizations of the random variables, which gives the quantity,  
$\{ \langle 0 | \rho^S | 1 \rangle \} $, 
where the curly brackets on $\langle 0 | \rho^S | 1 \rangle$ denote the average over the finite number of realizations.

\begin{figure}
	\begin{center}
  {\epsfxsize=3.5in\epsfysize=2.85in\centerline{\epsffile{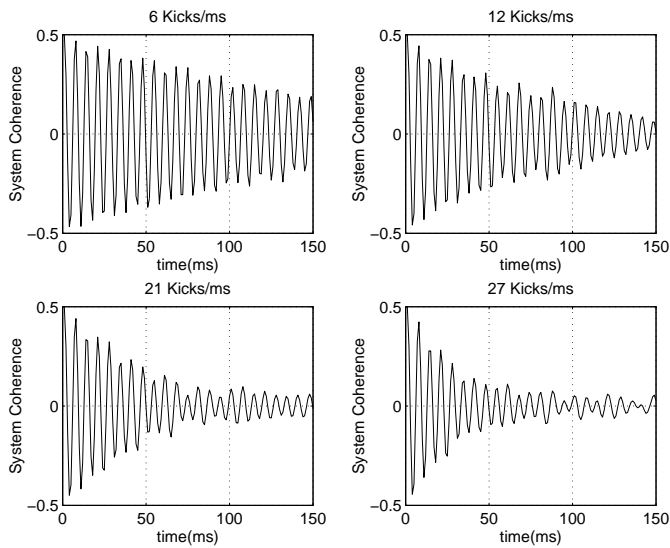}}}
\vspace{.25cm}
\caption{
Some example decays of the system coherence given by 
$\{ \mbra{0} \rho^S \mket{1} \}$ 
obtained from numerical simulation using Matlab for $50$
realizations of the random kick variables. 
The kick rates for each subplot are labeled above the figures. In this range 
the envelope of the decay is exponential (see inset to Fig.~\ref{DecoFig}).
We note that a higher kick rate leads to a faster system decay. \label{DecoFig2} }
	\end{center}
\end{figure}

\begin{figure}
	\begin{center}
  {\epsfxsize=3.5in\epsfysize=2.85in\centerline{\epsffile{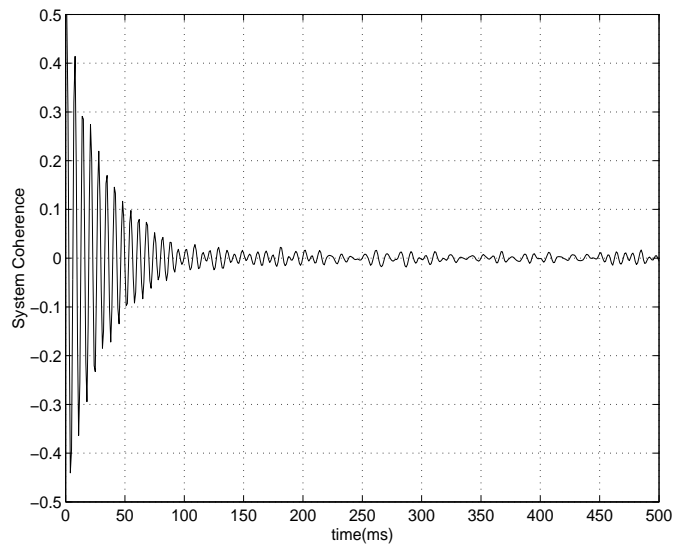}}}
\vspace{.25cm}
			\caption{A numerical simulation to demonstrate the suppression of revivals at longer times and higher averages. 
				The times go out to 
				500ms and the averages are taken for 200 realizations. Note that the revivals that seem
				prevalent in Figs.~\ref{DecoFig2} and ~\ref{DecoFig6} are diminished. \label{DecoFig8} }
	\end{center}
\end{figure}

\begin{figure}
	\begin{center}
\rotatebox{270}{{\epsfxsize=2.85in\epsfysize=3.5in\centerline{\epsffile{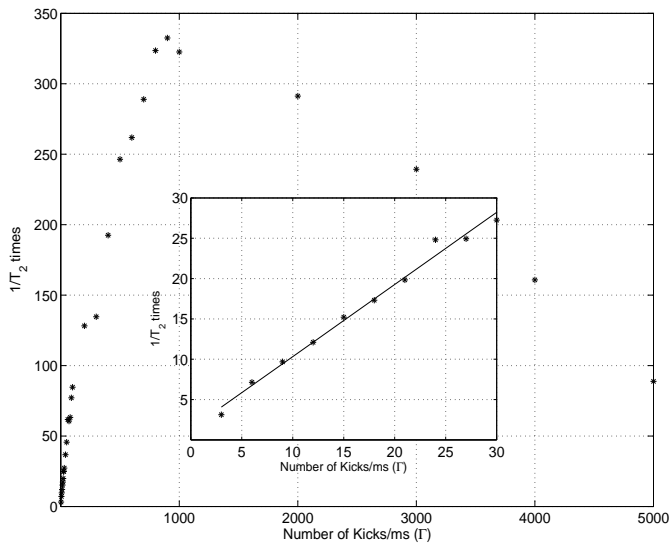}}}}
\vspace{.25cm}
			\caption{
Numerical simulation of the decoherence rate and the decoupling limit. 
Beyond a kick frequency of about $900$ kicks/ms the 
decoherence rate from the kicking starts to decrease. 
This transition to a decoupling effect is described in the text. 
After about $5000$ kicks/ms the decays are no longer exponential. 
Inset: Demsontration of the proportionality between decoherence rate and kick rate for low kick rates. 
This linear relationship can be understood from the analytic results obtained for 
the one-qubit environment. \label{DecoFig} }
	\end{center}
\end{figure}

In order to simulate the physical system used in the NMR study, 
we have selected the parameter values 
presented in
Table~\ref{Tab1}. The system and environment were initialized in the state $\sigma^S_xE^{E_1}_+E^{E_2}_+$ and we simulated
the evolution of the system on Matlab. We ran 10 different kick rates that ranged from $3$
kicks/ms to $30$ kicks/ms in steps of $3$. 
The kicks were sampled from a uniform distribution of angles
that ranged between $-\tfrac{\pi}{20}$ to $\tfrac{\pi}{20}$. The series was run for $150$ ms. 
We averaged over $50$ realizations
and obtain the plots shown in Fig.~\ref{DecoFig2}. 
As shown in Fig.~\ref{DecoFig8}, the late-time oscillations reflect the finite number of realizations of the random variables. 
The envelope of the decays in Fig.~\ref{DecoFig2} were fit to an exponential and the
decay constants exhibited a linear dependence on the kick rate for small kick rates, 
as expected from the analytic solution (see Fig.~\ref{DecoFig}). 
At about $900$ kicks/ms the decay rates start decreasing with increasing kick rate and the
system starts to become decoupled from the environment, an effect noted earlier in Eq.~\ref{f01g0}. 
This is the well-known decoupling phenomena in NMR \cite{Waugh}. 
The onset of decoupling occurs when the rms angle of the stochastic kicks 
approaches a rotation of $\pi$ (criticial damping). 
The rms angle is given by the typical kick size $\simeq \pi/10$ times 
the square root of the number of kicks over a cycle time $\simeq 1 /2J$ of the system-environment interaction.  
For the strongest system-environment coupling, $J \simeq 250$ Hz, the onset of decoupling 
is expected at a  kick rate of 800 kHz, in good agreement with the numerical results (see Fig.~\ref{DecoFig}). 

\section{IV. The NMR Implementation}

In this section, we describe the experimental implementation of our model. We chose propyne
\begin{figure}
	\begin{center}
  {\epsfxsize=3.2in\epsfysize=2.in\centerline{\epsffile{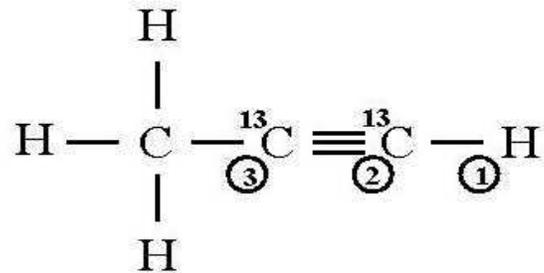}}}
\vspace{.25cm}
			\caption{The propyne molecule. The encircled labels on the $^{13}$Cs and the rightmost
hydrogen index the spins used in the experiment. The methyl group consists of the three hydrogens
and an unlabeled carbon. In the experiments the field of the spectrometer was $\sim9.2$ T and
the hydrogen resonances were $\sim400$ MHz while the carbon resonances were around $\sim100$
MHz. The chemical shift difference between the two labeled carbons is $1.260$ kHz. Using the indexing
scheme in the figure the $J$-coupling constants are as follows: $J_{12}=246.5$ Hz, $J_{23}=173.8$ Hz and
$J_{13}=51.8$ Hz. The longitudinal relaxation times are $T^1_1=8.7$s, $T^2_1=23$s and $T^3_1=43$s, while
the transverse relaxation times are $T^1_2=1.1$s, $T^2_2=1.9$s and $T^3_2=1.7$s. \label{DecoFig5} }
	\end{center}
\end{figure} 
as the physical system (see Fig.~\ref{DecoFig5} for the internal Hamiltonian parameters). 
The hydrogen indicated with a circled 1 represents the system qubit and the two  
carbons labeled with a circled 2 and a circled 3 represent the environment qubits $E_1$ and $E_2$, respectively. 
These spin-1/2 nuclei 
have a large resonance frequency offset, so the hydrogen and carbon can be addressed and
detected separately. The relatively large couplings present amongst these nuclei implies
the interactions take place over short times, and the long relaxation times allow one to observe the hydrogen signals over a
relatively long time span without significant natural decay. The experiments were carried out on a liquid solution of propyne 
using a Bruker Avance spectrometer. Neglecting the methyl group (because it couples in very weakly), the internal
Hamiltonian for propyne 
is given to a good approximation by
\beqnar 
{\cal H}_{int}=\pi[\nu_1\sigma^1_z+\nu_2\sigma^2_z+\nu_3
\sigma^3_z \hspace{2.5cm} \nonumber \\
\hspace{1.5cm}+\tfrac{1}{2}(J_{12}\sigma^1_z\sigma^2_z
+J_{23}\sigma^2\cdot\sigma^3+J_{13}\sigma^1_z\sigma^3_z)],
\label{ExpHam}
\eeqnar
where the $\nu$'s are Larmor frequencies and the $J$'s the spin-spin coupling constants in Hertz (the various values are
given in Fig.~\ref{DecoFig5}). Eq.~\ref{ExpHam} should be compared with Eq.~\ref{SysHam}. 
The non-secular coupling between the carbon nuclei 
can be observed in the carbon spectra but has a negligible effect on the relevant experimental results. 

A convenient choice for the initial state of system and environment is one where hydrogen is in a superposition state and
both carbons are in an eigenstate. By placing the methyl hydrogens in an eigenstate as well, they can be eliminated from
playing a role in the hydrogen spin dynamics. This was accomplished by using a highly selective rf pulse that irradiated a
spectral line corresponding to the state
\beqn
\sigma^H_xE^{C1}_+E^{C2}_+E^{M}_+,
\eeqn
where $E_+=\tfrac{1}{2}(I+\sigma_z)$, $H$ represents hydrogen, $C1$ carbon 1, $C2$ carbon 2 and $M$ the methyl hydrogens.
For this implementation we used a $5.5s$ EBURP1 \cite{Green,Freeman} pulse. The spectral resolution of this pulse was $.5$ Hz and
its design is such that it only generates a uniform excitation profile in the specified bandwidth. Ultimately, only $\sim1/10$th of
the maximum intensity was excited. Nonetheless, this yielded sufficient signal-to-noise ratio to carry out the experiments. 

The observed hydrogen signal corresponds to $\langle \sigma^H_x(t)+i\sigma^H_y(t) \rangle$, 
and is equivalent to tracing away the carbons. 
The peaks of hydrogen spectrum had linewidths of $\sim.4$ Hz.
Consequently, the hydrogen signals decayed very slowly and we were able to pick a $150$ ms portion of the absolute magnitude that
remained flat within one percent. 

The carbon spin dynamics consisted of a series of delays interleaved with pulses. During the delays the spins evolved under the internal
Hamiltonian. The pulse flip angles were randomly sampled from a uniform distribution that ranged between $-\pi/20$ to $\pi/20$ about the
$y-$axis. A cycletime of $1$ ms was defined within which the kick frequency ranged from $3$ kicks/cycle to $30$ kicks/cycle in steps of
$3$ for a total of $10$ different kick frequencies. The range of the kick frequency was limited by the shortest pulse the spectrometer
was capable of generating, which is $100$ ns. The time alloted for a sequence of one delay period followed by
a pulse was given by the cycletime/(number of kicks/cycle). Within this sequence the delay time is given by the total sequence time minus
the pulse-on time. The maximum pulse-on time was $10$ $\mu$s which corresponded to the maximum flip angle of $\pi/20$. The nutation
frequency for this RF field was $2500$Hz. (Compare this to the chemical shifts of the carbons which were separated by $1260$Hz). For a given
kick frequency, the length of the series of successive sequences of delay plus pulse, generated as described above, fit the total
acquistion time of $150$ ms. 

\begin{figure}
	\begin{center}
  {\epsfxsize=3.5in\epsfysize=2.85in\centerline{\epsffile{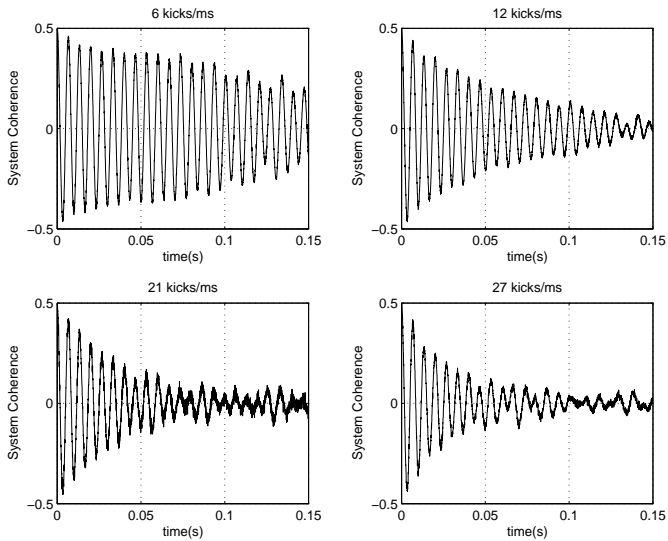}}}
\vspace{.25cm}
			\caption{Example decays from the experiment. The hydrogen signal was directly detected and the
				real part of the complex signal is plotted.  The fluctuations at the tail
				end of the higher kick rates are due to low statistics. This was confirmed by comparing with 
				simulations at higher averages. (See Fig.~\ref{DecoFig8}) \label{DecoFig6} }
	\end{center}
\end{figure}

\begin{figure}
	\begin{center}
  {\epsfxsize=3.5in\epsfysize=2.85in\centerline{\epsffile{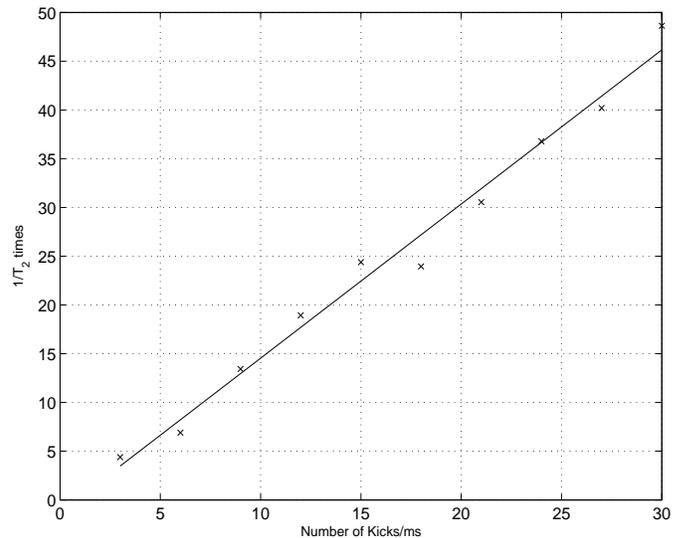}}}
\vspace{.25cm}
			\caption{ 
The linear dependence of the experimental decay constants on kick rate. 
The data point symbols (x) are larger than the 
error bars which range from $\pm.0075$ to $\pm.0573$. Compared to the slope
				in the simulations of Fig.~\ref{DecoFig} the experimental slope reflects faster decay.
				This disparity is due to the slight differences between the experiment and simulations. 
				\label{DecoFig7} }
	\end{center}
\end{figure}

The experiments were run for $10$ different kick frequencies with an average over $50$ realizations. A waiting time
of $300$s was used between successive realizations.
Fig.~\ref{DecoFig6} shows the result of the experiments. The absolute magnitude of these plots were fit to an exponential.
The $\chi^2$ per degree of freedom for the 10 fits ranged from $1.1$ to $8.5$. The $\chi^2$s fit the average decays well but 
don't account for the details in the fine structure evident from the oscillations of the magnitudes. As the kick frequency
increases the data demonstrates that the system is decohered faster. A plot of the decay constants as a function of kick
frequency, Fig.~\ref{DecoFig7}, shows this trend clearly. The experiment results seem to exhibit revivals in the higher
kick rates of Fig.~\ref{DecoFig7}. But this is due to low statistics (see Fig.~\ref{DecoFig8}).

\section{V. Discussion}

We have described a method for modeling decoherence that requires only limited quantum resources,  
and implemented the model on an NMR QIP. 
The key feature of the model which enables simulation of the 
dephasing effects and the attenutation of recurrences normally produced by 
a much larger quantum environment is the application of classical kicks to randomize 
the information in the environment states. 
Although 
the quantum system and environment dimensions are small and remain fixed, 
the system state exhibits an irreversible loss of coherence due to an averaging over the random 
realizations of kicks to the environment states. In particular, in the case of 
a $\sigma_z \sigma_z$ system-environment interaction 
we have shown that the kick frequency can be varied to control the decay rate of the phase-damping.  
Although in this paper we have focussed on the simulation of continuous phase-damping, 
the model can be immediately generalized to other system-environment couplings 
and the resultant decoherence channels. 
A major advantage of this model is that it provides a procedure through which the mechanisms of 
decoherence can be explored using techniques currently available in NMR QIP.


As resources permit, the model we have described may be generalized to simulate and study 
a wider variety of decoherence channels and system-environment couplings. 
In particular, the ``nearest'' quantum environment need not 
be the only quantum environment. For example, in order to implement a time-varying 
decoherence process with a fixed set of system-environment couplings 
it may be advantageous to introduce an environment ``hierarchy'' (see Fig.~1 for a schematic).    
The idea here is to couple the first quantum environment 
to a second, larger environment (through another set of 
fixed bilinear couplings), and so on.  
The dimension of the next Hilbert space in the 
environment hierarchy may be limited to $N^2$, where $N$ is the 
dimension of the Hilbert space of the immediately smaller system. 
In this framework only the nearest environment remains 
directly coupled to the system of interest. The approximation of   
using stochastic classical fields to reduce unwanted back-action 
may then be applied to the final quantum environment, which is 
much more remote from the system of interest. 

In conclusion, we have developed a model that is practical for simulating quantum decoherence effects 
associated with a time-independent superoperator on a QIP device. 
By varying the phase kicking rate in the stochastic Hamiltonian we can control the system's phase-damping rate.
In this presentation we have shown the
effectiveness of the methodolgy, in the case of one and  two spin environments, using analytical solutions, numerical 
simulations, and a physical implementation on an NMR QIP device. These methods have illustrated  the use of 
stochastic kick rates on the quantum environment for controlling system decoherence rates and recurrence times.

\section{Acknowledgements}
We thank L. Viola and E. Farhi for helpful discussions.
This work was supported by the U.S. Army Research Office under grant
number DAAD 19-01-1-0519 and DAAD 19-01-1-0678 from the Defense Advanced Research
Projects Agency. J.P.P. acknowledges support from Anpcyt, Ubacyt, and Fundacion Antorchas.

Correspondence should be addressed to DGC (email:{\it dcory@mit.edu}).


\begin{thebibliography}{99}

\bibitem{vonNeumannBook}
J. von Neumann, ``Measurement and reversibility''
and ``The measuring process'', chapters V and VI in
{\it Mathematische Grundlagen der Quantenmechanik}, Springer, Berlin, 1932;
English translation by R. T. Beyer
{\it Mathematical Foundations of Quantum Mechanics}, Princeton Univ. Press,
Princeton, 1955.

\bibitem{ZurekPT} Zurek, W. H., {\it Physics Today} {\bf 44}, 36 (1991).

\bibitem{GiuliniBook} Giulini, D., Joos, E., Kiefer, C., Kupsch, J.,
Stamatescu, I.-O., and Zeh, H. D., {\it Decoherence and the Appearance of
a Classical World in Quantum Theory}, Springer-Verlag, Berlin, 1996.

\bibitem{PazZurek} J. P. Paz and W. H. Zurek, 
``Environment induced superselection and the transition from 
quantum to classical'', in {\it Coherent matter waves, Les Houches 
Session LXXII} edited by R Kaiser, C Westbrook and F David, 
EDP Sciences, 533-614, Springer-Verlag, Berlin, 2001.

\bibitem{Zurek}
W. H. Zurek, {\em Phys. Rev. D} {\bf 26} 1862, 1982.

\bibitem{ZHP}
Zurek, W. H., Habib, S., and Paz, J. P., {\it Phys. Rev. Lett.}, 
{\bf 70}, 1187, (1993).

\bibitem{PZ99} Paz, J. P.,  and Zurek, W. H., {\it Phys. Rev. Lett.}
{\bf 82}, 5181 (1999).

\bibitem{Engeneering} J.E. Poyatos, I. Cirac and P. Zoller, 
{\it Phys. Rev. Lett.} {\bf 77} 4728-4731 (1997); A.R.R Carvalho,
P. Millman, R.L. de Mattos Filho and L. Davidovich, 
{\it Phys. Rev. Lett.} {\bf 86} 4988-4992 (2001), J.P. Paz,
{\it Nature} {\bf 412} 869-870 (2001). 

\bibitem{DecoRate} Paz, J. P.,  Habib, S., and Zurek, W. H.,
{\it Phys. Rev.} {\bf D 47}, 488 (1993); Anglin, J. R., Paz, J. P., 
and Zurek, W. H., {\it Phys. Rev.} {\bf A 53}, 4041 (1997).

\bibitem{Ekert}
G. M. Palma, K.-A. Suominen, and A. K. Ekert, Proc. R. Soc. London A {\bf 452}, 567 (1996).

\bibitem{Boson}
L. Viola and S. Lloyd, {\em Phys. Rev. A} {\bf 58}, 2733, 1998.

\bibitem{Legget}
Legget, A. J., Chakravarty, S., Dorsey, A. T., Fisher, M. P. A., Garg, A., and Zwerger, W.,
{\em Rev. Mod. Phys.} {\bf 59}, 1 (1987).

\bibitem{Knill-QEC}
E.~Knill and R.~Laflamme, {\em Physical Review A}, {\bf55} 900--911, 1997.

\bibitem{Knill-NS}
E.~Knill, R.~Laflamme, and L.~Viola, {\em Phys. Rev. Lett.}, 84:2525, 2000.

\bibitem{Viola-NS}
L.~Viola, E.M. Fortunato, M.A. Pravia, E.~Knill, R.~Laflamme, and D.G. Cory, {\em Science}, 293:2059--63, 2001.

\bibitem{Fortunato-DFS}
E.M. Fortunato, L.~Viola, J.~Hodges, G.~Teklemariam, and D.G. Cory, {\em New Journal of Physics}, 4:5.1--20, 2002.

\bibitem{CC-T}
C.~Cohen-Tannoudji, J. Dupont-Roc, and G. Grynberg, {\it Atom-Photon Interactions: Basic Processes and Applications},
J. Wiley , NY, 1992.

\bibitem{Kraus}
K. Kraus, {\em States, Effects and Operations: Fundamental Notions of Quantum Theory}, Springer-Verlag, Berlin, 1983.

\bibitem{NielsenBook}
M.A. Nielsen and I.L. Chuang, {\em Quantum Computation and Quantum Information}, Chapter 8,
Cambridge University Press, {Cambridge, UK}, 2000.

\bibitem{Somaroo}
S. S. Somaroo, C. H. Tseng, T. F. Havel, R. Laflamme, and D. G. Cory, {\em Phys. Rev. Lett.}, 82:5381, 1999.

\bibitem{Tseng}
C. H. Tseng, S. S. Somaroo, Y. Sharf, E. Knill, R, LaFlamme, T. F. Havel, and D. G. Cory,
{\em Phys. Rev A}, A {\bf 61} 012302 (2000).

\bibitem{Waugh}
J. S. Waugh, {\it J. Mag. Res.} {\bf 50}, 30-49 (1982).

\bibitem{PreskillNotes}
J.~Preskill, {\em Lecture notes on quantum computation},
Chapter 3, http://www.theory.caltech.edu/\~{}preskill/ph219.

\bibitem{Schumacher}
B.~Schumacher, \newblock {\em Physical Review A}, 54:2614--28, 1996.

\bibitem{Havel}
T. F. Havel, Y. Sharf, L. Viola, and D. G. Cory, {\em Phys. Lett. A}, {\bf 280}, 282 (2001).

\bibitem{DiVincenzo}
DiVincenzo, D. P., {\em Fort. der Phys.} {\bf 48}, 771-83, (2000).

\bibitem{Green}
H.~Green and R.~Freeman, {\em Journal of Magnetic Resonance}, 93:93--141, 1991.

\bibitem{Freeman}
R. Freeman, {\em Spin Choreography}, Oxford Univ. Press, Oxford, UK (1998).

\end{thebibliography}
\end{document}